# An AI-Driven Approach to Wind Turbine Bearing Fault Diagnosis from Acoustic Signals

Zhao Wang,[1] Xiaomeng Li,[1] Na Li,[2] Longlong Shu [3,a)]
[1]*School of Computer and Network Engineering, Shanxi Datong University, Datong 037009, China*
[2]*R&D Centre, CRRC Yongji Electric Co., Ltd, Xi'an 710018, China*
[3]*Nanchang University, Nanchang 330031, China*

Abstract: This study aimed to develop a deep learning model for the classification of bearing faults in wind turbine generators from acoustic signals. A convolutional LSTM model was successfully constructed and trained by using audio data from five predefined fault types for both training and validation. To create the dataset, raw audio signal data was collected and processed in frames to capture time and frequency domain information. The model exhibited outstanding accuracy on training samples and demonstrated excellent generalization ability during validation, indicating its proficiency of generalization capability. On the test samples, the model achieved remarkable classification performance, with an overall accuracy exceeding 99.5%, and a false positive rate of less than 1% for normal status. The findings of this study provide essential support for the diagnosis and maintenance of bearing faults in wind turbine generators, with the potential to enhance the reliability and efficiency of wind power generation.

## Introduction

Wind power is a rapidly growing electricity generation technology within the realm of low-carbon energy. Its development is a major strategy for carbon emission reduction around world, and it has achieved significant progress in recent years. However, this growth has brought about some urgent issues that need to be addressed. With the widespread adoption of wind power, the number of wind power devices has increased globally. Wind power is now receiving greater attention from various countries worldwide. Nevertheless, the long-term maintenance and upkeep of wind power equipment have significantly impacted the economic viability of wind power projects. To remain competitive in modern wind power generation, it is imperative to minimize the risk of failures, reduce maintenance costs, and enhance the availability and efficiency of the system[1-3].

The wind turbine generator, as the core component of a wind power system, can lead to substantial maintenance costs for manufacturers in the event of a malfunction. While the structure of wind turbine generators is complex, the drivetrain is the most susceptible component to failures during operation[4-6]. In real-world fault detection scenarios, issues with the drivetrain can often be revealed through noise. The human auditory system is a unique nonlinear system with varying sensitivities to different frequency signals. Experienced engineers can identify the location and even the type of faults by listening to the noise. However, the traditional method of relying on human ears for noise identification in maintenance is inefficient, highly subjective, lacks uniform fault determination standards, and heavily relies on expert experience.

Therefore, there is a need to develop an acoustic fault diagnosis method that can automatically extract useful features from noise and provide self-learning capabilities, all through artificial intelligence techniques. Artificial intelligence is an emerging technology supported by machine learning and deep learning algorithms, which can provide sustainable technical support for the wind power industry[7-12].

In the field of audio recognition, one of the most commonly used features is Mel-frequency cepstral coefficients (MFCC)[13]. MFCC is a feature that accurately describes the envelope of the short-term power spectrum of an audio signal. It takes into consideration human auditory characteristics by mapping the linear frequency spectrum into a Mel-scale nonlinear frequency spectrum and then converting it into cepstral coefficients. After further processing, this feature can be used as input for audio analysis. The process of extracting MFCC features typically involves the following steps: a) Audio segmentation and windowing; b) Short-time analysis window and FFT transformation; c) Mel-frequency spectrum filtering; d) Mel-frequency cepstral analysis (Discrete Cosine Transform). However, in the field of bearing faults diagnosis applications, when engineers have limited expertise in both audio feature analysis and bearing domain knowledge, or when there are numerous types of bearing faults and significant noise interference, the classification performance can be compromised. Low classification accuracy, potential misclassification, poor generalization capabilities, making it challenging to achieve precise classification and diagnosis of motors under varying operating conditions, such as different rotation speeds. In order to address these challenges and improve the classification and diagnosis of bearing faults in wind turbines, more advanced techniques, including deep learning approaches, can be considered. Deep learning models, such as convolutional neural networks (CNNs) and recurrent neural networks (RNNs), are capable of capturing both linear and nonlinear features and can automatically learn complex patterns and representations[14-18]. These models have shown promising results and can potentially enhance the accuracy and generalization of fault classification in wind turbine bearings.

To fulfill diagnosis of bearing faults in various audio recognition tasks, we need combine the deep learning and audio feature analysis. However, it's crucial to have a sufficiently large and diverse dataset for training deep learning models and to carefully fine-tune the model architecture and hyperparameters to suit the specific task. Additionally, domain knowledge about bearing faults and wind turbine operations can still be valuable for feature engineering and model interpretation, even when using deep learning techniques. In this paper, a combination of frequency domain feature extraction and deep learning models was employed to achieve the classification of bearing fault audio signals.

## Experiment Details

The method for constructing a classification model for wind turbine generator bearing faults is achieved through the following steps:

*Step 1: Predefined Bearing Fault Types and Quantity (5 Types)*

Fault Type 1: Both drive-side and non-drive-side bearings are normal.
Fault Type 2: Drive-side bearing with inner race spalling, Non-drive-side bearing is normal.
Fault Type 3: Drive-side bearing is normal, Non-drive-side bearing with inner race spalling.
Fault Type 4: Drive-side bearing with outer race spalling, Non-drive-side bearing is normal.
Fault Type 5: Drive-side bearing is normal, Non-drive-side bearing with outer race spalling.

*Step 2: Acquisition and Framing of Raw Signals*

The wind turbine generator is mounted on a test bench with the rotor short-circuited, operating at an unloaded speed of 2000 rpm. The audio acquisition device features five audio inputs with 16-bit precision and a sampling frequency of 48 kHz. For each type of bearing fault, continuous audio data is collected for 40 seconds at positions ranging from 0.5 meters to 1.5 meters from the non-drive side of the wind turbine generator. This results in 25 sets of original audio data, with each original signal segmented into 400 frames. Each frame, with a time interval of t equal to 0.1 seconds, is labeled with its respective fault type. This process generates an original signal dataset comprising 400 time-series data sequences, each tagged with fault labels.

The audio acquisition device's sampling frequency is 48 kHz, yielding 4800 samples for each time-series data sequence. We apply Fast Fourier Transform (FFT) to each of the 4800 samples using the FFT algorithm, resulting in FFT values for each sample point. This integrates time-domain values with their corresponding frequency-domain values, transforming each time-series data sequence from one-dimensional in the time domain (1x4800) to two-dimensional, encompassing both time and frequency domains (2x4800).

Subsequently, the data is randomly divided into training, validation, and test sets in an 8:1:1 ratio. Each original signal dataset provides 320 (400x80%) training data, 40 (400x10%) validation data, and 40 (400x10%) test data. Ultimately, all 25 original datasets collectively yield 8000 (25x320) training samples, 1000 (25x40) validation samples, with the remaining 1000 (25x40) data points serving as test samples.

*Step 3: Establishing a Deep Learning Network*

Using the training samples from Step 2, a convolutional LSTM model was constructed consisting of 17 layers, as illustrated in Figure 3. To input sequences to the network, use a sequence input layer. To use convolutional layers to extract features, that is, to apply the convolutional operations to each frame of the audio signals independently, use a sequence folding layer followed by the convolutional layers, and then a sequence unfolding layer. To use the LSTM layers to learn from sequences of vectors, use a flatten layer followed by the LSTM and output layers. The detailed parameters of the neural network structure are recorded in Table 1.

The training data from Step 2 was used for network training, with specific training parameters set as follows: The model was trained for 10 epochs with mini-batches of size 128, employing the 'SGD' solver and specifying a learning rate of 0.01. To assess the recognition accuracy and generalization ability of the training network, we utilized the validation samples from Step 2. As depicted in Figure 4, throughout network training, both the training and validation sample loss functions continued to decrease, while accuracy consistently increased. The accuracy reached 99% or higher, leading us to save the model at this point. The validation accuracy served as a convergence criterion for the network.

*Step 4: Validating Model Classification Results*

The test samples from Step 2 are input into the model created in Step 3 for classification. Subsequently, the final results for the five categories are validated, attaining an overall accuracy surpassing 99.5%. Additionally, the false positive rate for normal operating states remains below 1%. The results displayed high accuracy during training on the samples and demonstrated excellent generalization ability during validation. This indicates that the model performs well in terms of generalization capability.

## Discussions

The primary contribution of this research lies in the proposal and implementation of a convolutional LSTM model for wind turbine bearing fault classification. In comparison to the diagnosis models using signal from the accelerometer sensors, the non-accelerometer way could effectively reduce the hardware costs associated with data acquisition and bring substantial economic benefits, which exhibits significant enhancements in multiple dimensions. Furthermore, it leverages the strengths of both CNN and LSTM, allowing it to capture temporal information and spatial features concurrently. This innovative design holds considerable potential for bearing fault classification tasks, which represents a noteworthy advancement, surpassing previous research results in the field of bearing fault classification. Encouragingly, the model exhibits robust generalization across diverse operational conditions of wind turbines, a critical factor for real-world applications in wind farms. This indicates that our model excels not only on standard test sets but also maintains its performance when confronted with new operating conditions and unknown fault types. This holds paramount importance for enhancing wind turbine reliability and reducing maintenance costs. Our research also underscores the pivotal role of data preprocessing. Through meticulous data collection and preprocessing methods, it is possible to significantly enhance the model's effectiveness, which carries implications for future research and practical applications.

Despite the substantial progress in our research, there remain certain limitations to address. For instance, further refinements may be necessary to adapt our model for fault classification under extreme environmental conditions. Additionally, future research can explore the integration of additional sensor data into the model to improve classification performance. Finally, we emphasize the practical applications and societal impact of our research within the wind energy sector. By enhancing the efficiency of wind turbine fault diagnosis, we can contribute to the sustainable development of renewable energy and reduce maintenance costs.

# Conclusions

In this paper, we begin by collecting audio data from bearings with various types of faults, carefully labeling audio segments corresponding to each fault type. Subsequently, we employ a combination of frequency domain feature extraction and a convolutional LSTM model to classify the audio signals related to bearing faults. This approach harnesses a substantial amount of labeled data to train deep learning models capable of learning and recognizing the acoustic features specific to different fault types. This, in turn, enables accurate classification and fault diagnosis. Furthermore, through the incorporation of frequency domain feature extraction methods, we effectively capture vital information within the audio data, thereby enhancing model performance and generalization capacity. This holistic approach holds significant potential for advancing the field of bearing fault detection and diagnosis.

## Figure captions:

Figure 1.

Picture of bearing preset faults, including normal and spalling.

Figure 2.

(a) Time-domain waveform graphs of various fault types.

(b) Time-frequency spectral graphs of various fault types.

Figure 3.

Schematic diagram of the convolutional LSTM architecture.

Figure 4.

Training process diagram of the network.

Figure 5.

Confusion matrix of the test dataset for evaluating the performance of the model.

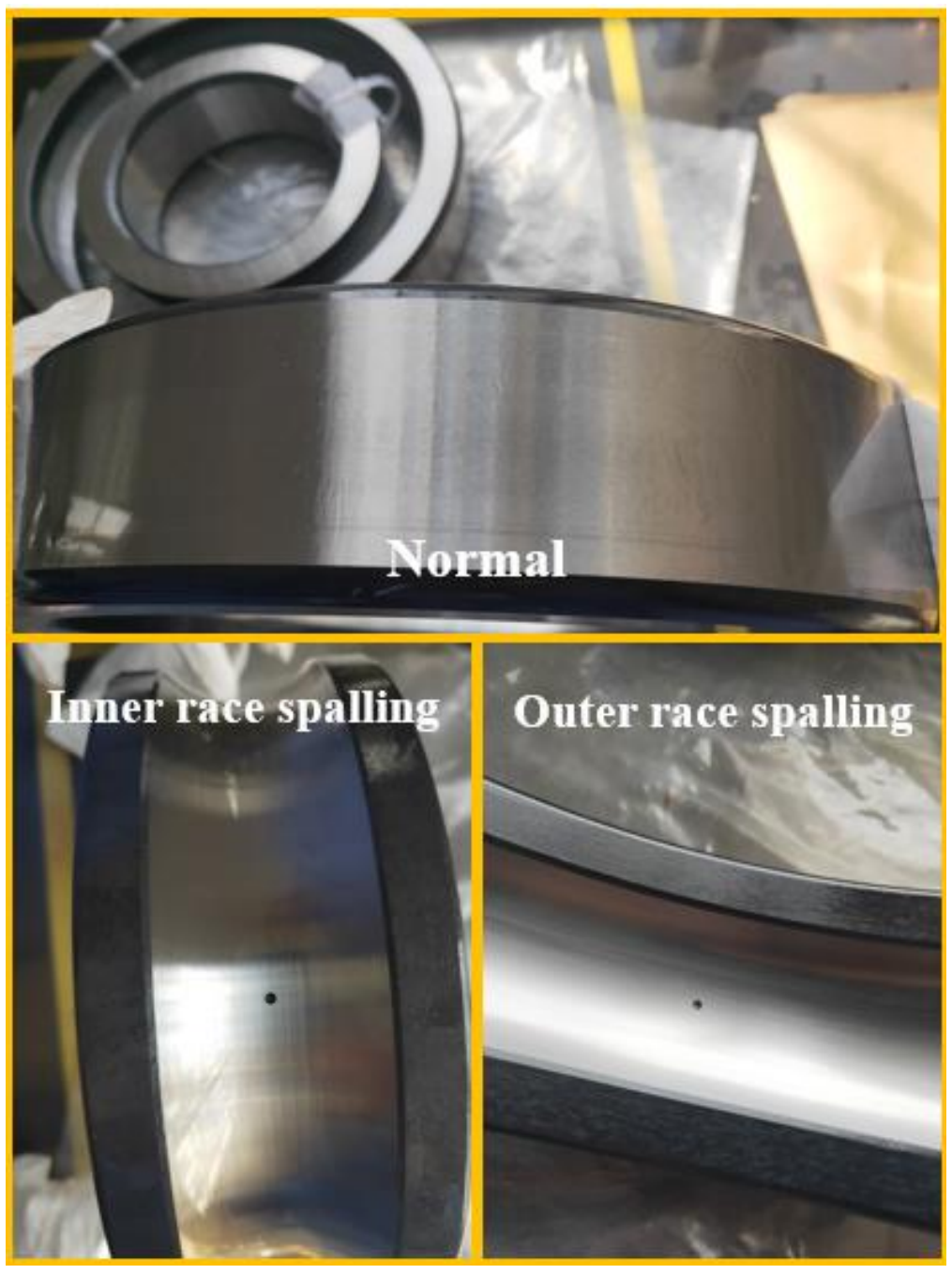

Figure 1

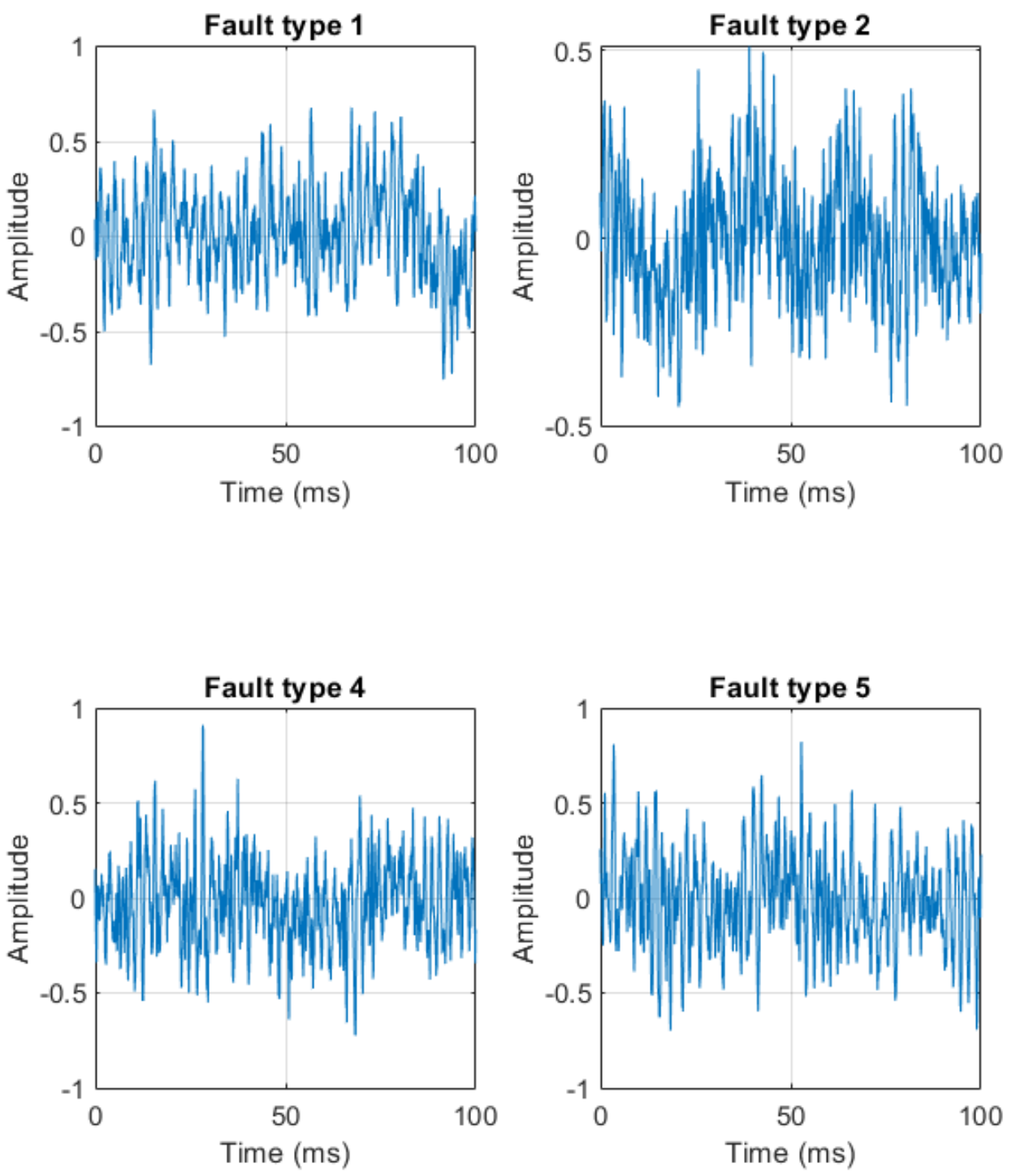

Figure 2a

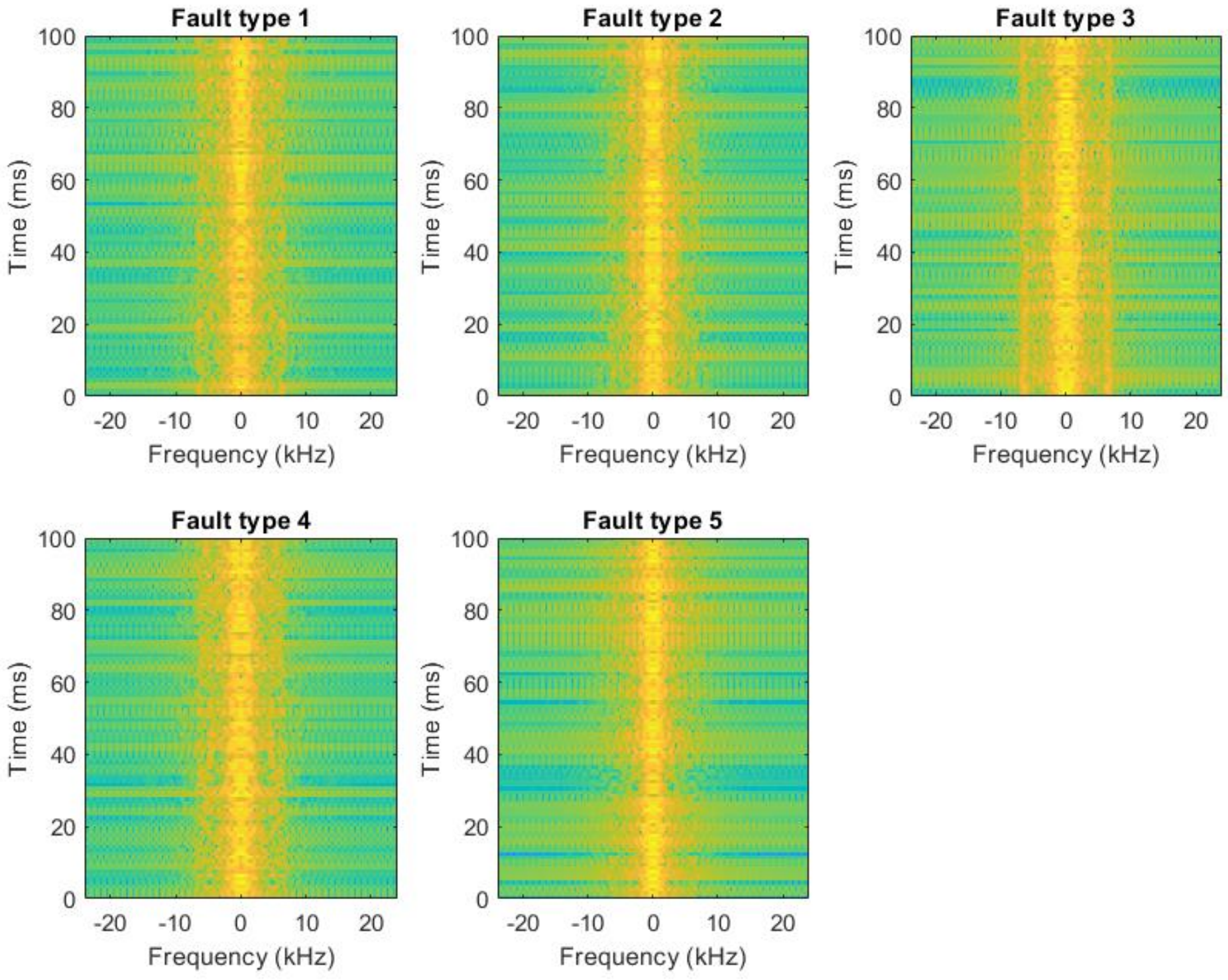

Figure 2b

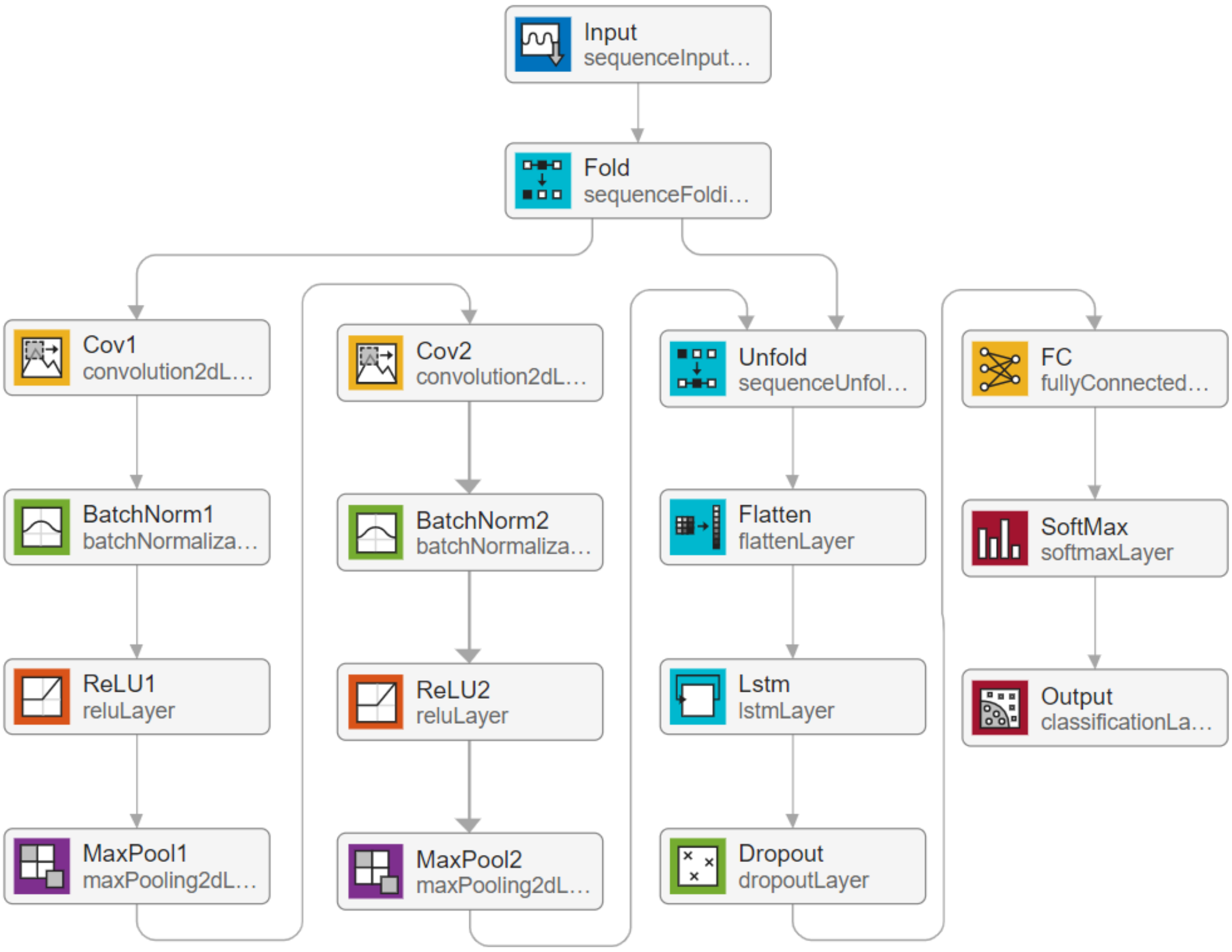

Figure 3

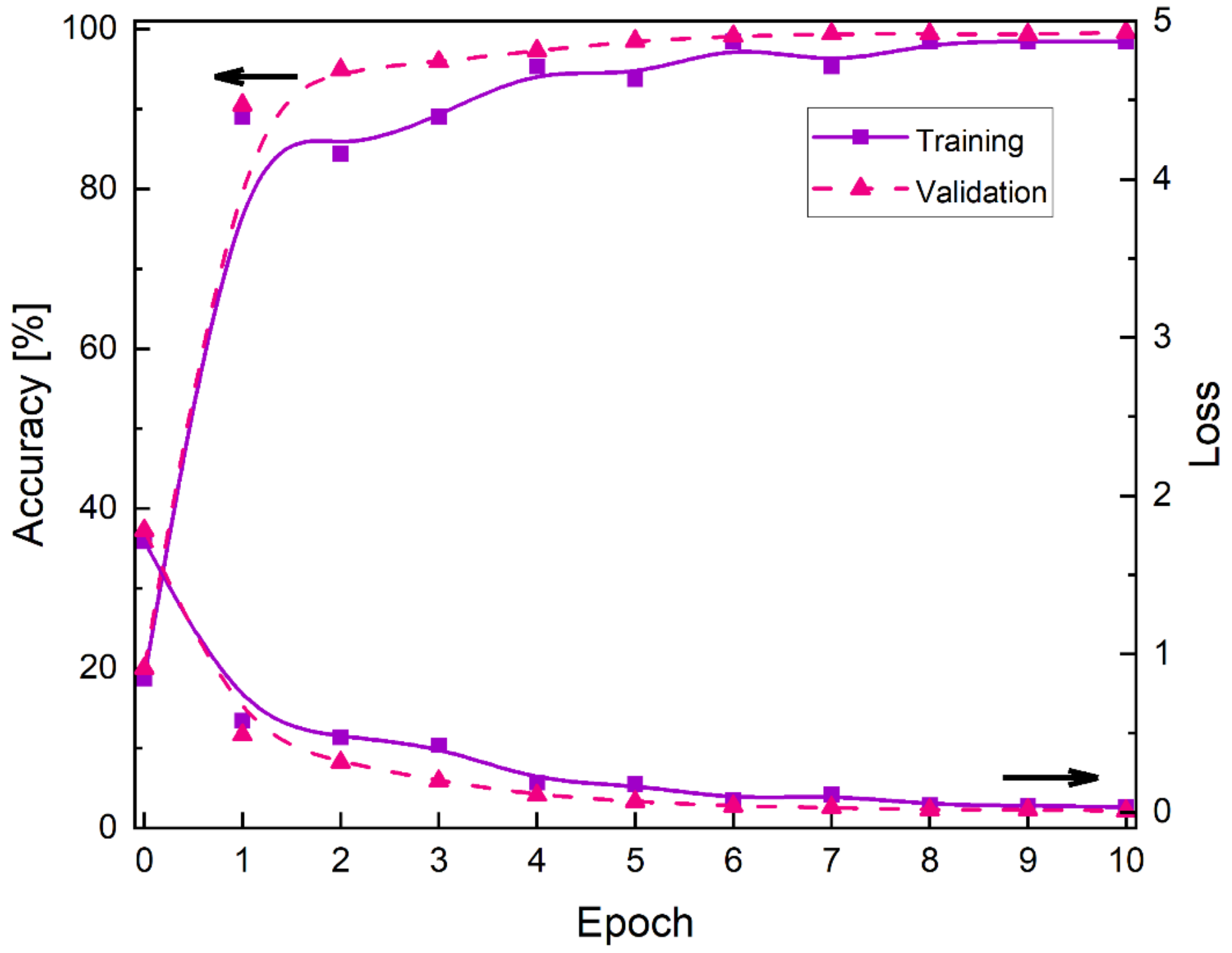

Figure 4

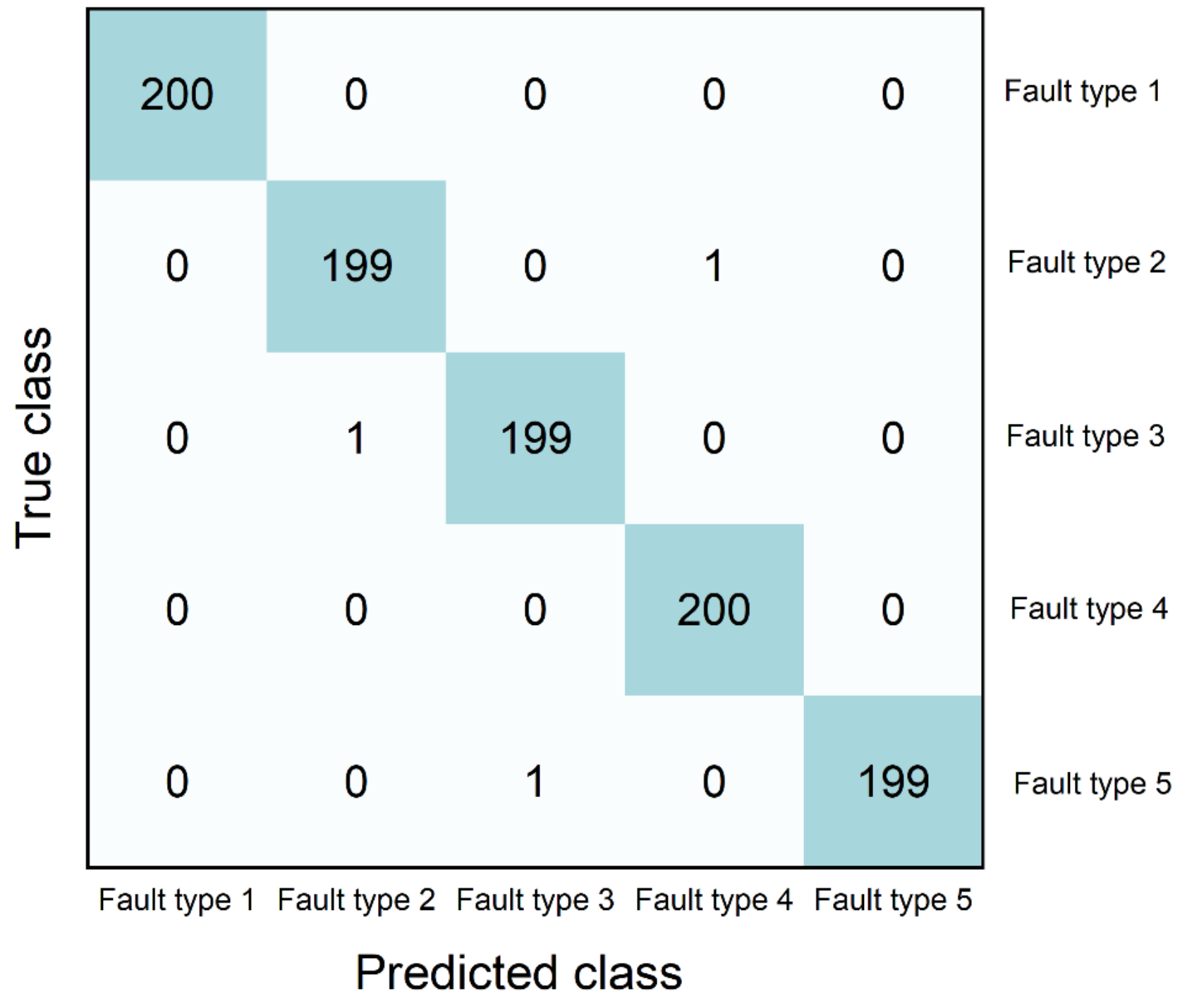

True class
200 0 0 0 0 Fault type 1
0 199 0 1 0 Fault type 2
0 1 199 0 0 Fault type 3
0 0 0 200 0 Fault type 4
0 0 1 0 199 Fault type 5
Fault type 1 Fault type 2 Fault type 3 Fault type 4 Fault type 5
Predicted class

Figure 5

Table 1

| Layers | Details | Activations | Learnables |
| --- | --- | --- | --- |
| *Input* | *Sequence input* | *2×4800×1* | - |
| *Fold* | *Sequence folding* | *out 2×4800×1*<br>*MiniBatchSize 1* | - |
| *Cov1* | *16 1×8 convolutions with stride [1 1] and padding 'same'* | *2×4800×16* | *Weights 1×8×1×16*<br>*Bias 1×1×16* |
| *BatchNorm1* | *Batch normalization* | *2×4800×16* | *Offset 1×1×16*<br>*Scale 1×1×16* |
| *ReLU1* | *ReLU* | *2×4800×16* | - |
| *MaxPool1* | *1×2 max pooling with stride [1 2]* | *2×2400×16* | - |
| *Cov2* | *24 1×8 convolutions with stride [1 1] and padding 'same'* | *2×2400×24* | *Weights 1×8×16×24*<br>*Bias 1×1×24* |
| *BatchNorm2* | *Batch normalization* | *2×2400×24* | *Offset 1×1×24*<br>*Scale 1×1×24* |
| *ReLU2* | *ReLU* | *2×2400×24* | - |
| *MaxPool2* | *1×2 max pooling with stride [1 2]* | *2×1200×24* | - |
| *Unfold* | *Sequence unfolding* | *2×1200×24* | - |
| *Flatten* | *Flatten* | *57600* | - |
| *Lstm* | *LSTM with 100 hidden units* | *100* | *InputWeights 400×57600*<br>*RecurrentWeights 400×100*<br>*Bias 400×1* |
| *Dropout* | *50% dropout* | *100* | - |
| *FC* | *5 fully connected layer* | *5* | *Weights 5×100*<br>*Bias 5×1* |
| *Softmax* | *softmax* | *5* | - |
| Output | crossentropyex | 5 | - |